\documentclass[twocolumn,           
               showpacs,            
               preprintnumbers,     
               aps,                 
               prd,          	    
               a4paper,             
               superscriptaddress,  
               unsortedaddress,     
               footinbib,           
               tightenlines,        
               floats               
               ]{revtex4}

\usepackage{graphicx}
\usepackage{latexsym}
\usepackage{amsmath,amssymb}        
\usepackage[draft=false]{hyperref}

\newcommand{\al}{\ensuremath{\alpha}}
\newcommand{\ga}{\ensuremath{\gamma}}

\newcommand{\ep}{\ensuremath{\epsilon}}

\newcommand{\om}{\ensuremath{\omega}}

\newcommand{\p}{\ensuremath{\phi}}
\newcommand{\vp}{\ensuremath{\varphi}}

\renewcommand{\L}{\ensuremath{{\cal L}}}

\newcommand{\ra}{\ensuremath{\rightarrow}}
\newcommand{\del}{\ensuremath{\partial}}
\newcommand{\Del}{\ensuremath{\nabla}}

\newcommand{\half}{\ensuremath{\frac{1}{2}}}

\newcommand{\be}{\begin{equation}}
\newcommand{\ee}{\end{equation}}
\newcommand{\ba}{\begin{eqnarray}}
\newcommand{\ea}{\end{eqnarray}}

\begin{document}


\title{Dynamics of Q-Balls in an expanding universe}
\author{Eran Palti}
\affiliation{Department of Physics and Astronomy, University of Sussex, 
             Brighton BN1 9QJ, United Kingdom}
\author{P. M. Saffin.}        
\affiliation{Department of Physics and Astronomy, University of Sussex, 
             Brighton BN1 9QJ, United Kingdom}
\author{E.J. Copeland}
\affiliation{Department of Physics and Astronomy, University of Sussex, 
             Brighton BN1 9QJ, United Kingdom}
\pacs{98.80.Cq \hfill hep-th/xxxxxxx}
\preprint{hep-th/xxxxxxx}
\date{\today} 


\begin{abstract}

We analyse the evolution of light Q-balls in a cosmological background, and find a number of interesting features. 
For Q-balls formed with a size comparable to 
the Hubble radius, we demonstrate that there is no charge radiation, 
 and that the Q-ball maintains a constant physical radius. Large expansion rates cause
charge migration to the surface of the Q-ball, corresponding to a non-homogeneous internal rotation
frequency. We argue that this is an important phenomenon as it leads to a large surface 
charge and possible fragmentation of the Q-ball. We also explore the deviation of the Q-ball profile 
function from the static case. By introducing a parameter $\epsilon$, which is the ratio 
of the Hubble parameter to the frequency of oscillation of the Q-ball field, and using solutions to 
an analytically approximated equation for the profile function, we determine the dependence of the new features
on the expansion rate. This allows us to gain an understanding of when they should be considered and when they can 
be neglected, 
thereby placing restrictions on the existence of homogeneous Q-balls in expanding backgrounds.  

\end{abstract}

\maketitle


\section{Introduction}

Q-balls \cite{kn:coleman} are non-topological solitons that exist as minimum energy configurations 
of a complex scalar field (typically a flat direction in a MSSM model 
\cite{kn:kusenko,kn:shaposhnikov,kn:mcdonald}) with a global U(1) symmetry;
the gauged version has been studied in \cite{kn:lee}. First 
introduced by Coleman \cite{kn:coleman}, following pioneering work by Friedberg, Lee and Sirlin
\cite{kn:friedberg}, their formation, particularly through an 
Affleck-Dine condensate collapse 
\cite{khlebnikov,kn:enqvist,kn:kawasaki}, has been 
frequently investigated . However, little work has gone into looking at the behaviour of general 
fully formed Q-balls in a cosmological background. The need to do this is  becoming increasingly 
important as Q-balls are often discussed in a cosmological context, typically as a 
possible catalyst for baryogenesis \cite{kn:mcdonald} and as candidates for dark 
matter \cite{kn:steinhardt,kn:shaposhnikov}. In this short paper we begin this investigation by considering    
light Q-balls (where the Q-ball energy density is dominated by the background 
energy density) in a cosmological setting.

The background expansion leads to changes in the Q-ball behaviour from the static case. We identify a parameter 
$\epsilon$, which is the ratio
of the Q-ball oscillation frequency (in the core of the Q-ball) and the Hubble parameter, 
as a measure of the deviation from static background solutions. We then 
show that in an expanding background the Q-ball characteristics are determined by $\epsilon$. By  
numerically obtaining the explicit relation between $\epsilon$ and the deviations from the static solutions, 
we are able to
 answer two important questions: when should we be considering background expansion effects and how do these 
effects depend on the
 rate of expansion? Turning explicitly to the effect of the expansion on the Q-ball we show 
that the Q-ball parameters exhibit expansion driven oscillations and that the expansion causes the field rotation 
frequency to become non-homogeneous, driving the conserved charge to the surface of the Q-ball but not radiating 
it, whilst maintaining a constant physical radius. 
In what follows, we begin by summarising the basics of Q-balls in flat space leading to a 
discussion of the problems a dynamic background introduces and our proposed 
way of analysing the new solutions. In section \ref{numerical} we numerically solve the full system 
of a light Q-ball in an 
expanding Universe,  and 
explicitly demonstrate the consistency of introducing the parameter $\epsilon$. We introduce
our analytic approach in section \ref{analytic} 
and discuss its uses and implications in section \ref{implications}.
We summarise our results in section \ref{conc}.

\section{Q-balls in an expanding background} \label{dynamic}
The evolution of a complex scalar field $\vp$ with a Lagrangian density $\L = 
\left|\del\vp\right|^2  - V(|\vp|^2)$ is given by 
\be
\Box\vp = \frac{\del V}{\del\vp^*}.
\ee
For a spherically symmetric field of the form \mbox{$\vp=e^{i\om t}f(r)$}, where the frequency $\om$ and 
profile function $f$ are
real, in a flat background we have
\be
\Del^2f + \frac{\del V_{eff}}{\del f} = 0, \label{flateqnofmotion}
\ee
where $V_{eff} = \half\om^2f^2 - \half V(f^2)$. Equation (\ref{flateqnofmotion}) 
is equivalent to Newtonian motion, with a time dependent friction term, of a particle in a 
potential $V_{eff}$, with r acting as time. A Q-ball corresponds to a solution where the particle 
starts at some point $\vp_0\equiv \vp(r=0)$ and rolls down the potential coming to 
rest at the origin. This would be a field configuration with $\vp\ra 0$ as 
$r\ra\infty$, a necessary condition for finite energy, if we require $V(0)=0$.
The conditions for such a solution to exist are 
that $V_{eff}$ is 
decreasing near the origin (allowing the particle to slow down and come to rest 
there) and that at the starting point 
$V_{eff}\left(\left|\vp_0\right|^2\right)\geq 0$, so the particle has enough energy to reach $\vp=0$.
Hence for a flat space Q-ball solution to exist, the value of  $\om$ is restricted to be in the range      
\be
\frac{2V\left(\left|\vp_0\right|^2\right)}{\left|\vp_0\right|^2} \leq \om^2 \leq 
V''(0). \label{omegaconstraints}
\ee
Assuming appropriate values of $\om$ are found, then equation (\ref{flateqnofmotion})
has an exact soliton (Q-ball) solution with a stabilising (against decay) Noether 
charge given by \cite{kn:coleman}
\be
Q=\om\int|\vp|^2 d^3x.
\ee
In a Robertson-Walker universe with scale factor $a(t)$, the equation of motion becomes
\be
\frac{1}{a^2}\left(\Del^2\vp\right) - \frac{\del^2\vp}{\del t^2} - 
\frac{3}{a}\left(\frac{\del a}{\del t}\right)\left(\frac{\del\vp}{\del t}\right)
=\frac{\del V}{\del\vp^*}.   \label{eqofmotintime}
\ee
In this case the ansatz $\vp=e^{i\om t}f(r)$ with constant $\om$ is no longer a solution 
and so the 
flat space Q-ball scenario of a spherically symmetric field configuration with a 
constant internal rotation frequency does not hold. Now, in most cosmological 
settings we may neglect the Hubble drag term as a first approximation, however a 
better and more instructive approach is to keep all the terms and simply work in conformal time $\eta$ 
defined by $dt=a(\eta)d\eta$. Introducing $\p=\vp 
a(\eta)$ we obtain the following evolution equation 
\be
\Del^2\p - \p'' = a^4\frac{\del}{\del\p^*}V(|\phi|^2/a^2) - \left(\frac{a''}{a}\right)\p 
\label{expandingeqnofmotion}
\ee
where time derivatives with respect to $\eta$ are denoted by a prime. The utility of this
approach is that all the time dependence of the background has been re-written as a time
dependent potential for $\phi$, 
\mbox{${\cal V}(|\phi|^2)=a^4V(|\phi|^2/a^2)-(a''/a)|\p|^2$}.
\section{Numerical simulations} \label{numerical}    

For our analytic approach in section \ref{analytic} we introduce a small parameter 
$\ep=H/\omega_0$, so
for our purposes a natural background for looking at the evolution of the Q-ball profile is 
de-Sitter space. In this case the parameter $\ep$ is a constant and so 
by looking at the profile evolution over a constant range of the scale-factor (constant number of efoldings), but at 
different rates of expansion, we can explore the dependence of the approximation 
on $\ep$. 
First of all though we shall perform simulations of the full evolution, (\ref{expandingeqnofmotion}),
so we can later test our analytic method.
In our simulations we used a form of the potential motivated from supersymmetry, namely 
\be
V(\vp)=\alpha \vp^2 - \beta \vp^4 + \gamma \vp^6 \label{potential}
\ee
with $\alpha=2.3 \times 10^7 {\rm GeV}^2$, $\beta=1.0$, and $\gamma=1.0 \times 10^{-8} {\rm GeV}^{-2}$. 
The chosen values correspond to the small field expansion of a SUSY potential of the form 
$V(\vp) = M^4 \ln (1 + \vp^2/M^2)$ where $M \sim 10^4$ GEV is the scale of SUSY breaking. 
Recall that we are considering light Q-balls, defined as those whose dynamics are determined by 
the background spacetime, and whose own contribution to the energy density in the Friedmann equation 
is negligible. Given that, we choose an appropriately large value $H \sim 10^{2}$ GeV for the Hubble 
parameter. This corresponds to a background energy scale of $10^{10}$ GeV, or an evolution in the 
radiation dominated era, just after say reheating at the end of a period of early Universe inflation. 
In order to test the robustness of the parameter $\epsilon$ in determining the true solution, we ran 
a series of simulations in which we  fixed the charge of the Q-ball and varied $H$, increasing it 
from a starting value of $100$GeV. We verified our results for
a number of values of Q but the results we show here correspond to Q=300, (giving $\ep=0.05$ for $H=100$GeV).

\begin{figure}[t]
\includegraphics[scale=0.38,angle=0]{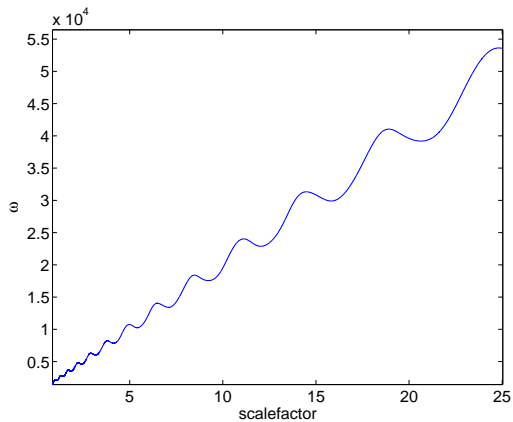}
\caption{Figure showing $\om$ in the core of the Q-ball as a function of the 
scalefactor.}
\label{omega}
\end{figure}

Starting with the profile of a static configuration, we obtained the evolution of 
the Q-ball as we turned on the expansion
of the universe, reaching a de-Sitter background by $a=1$. We then evolved the profile with this 
background and a fixed value of $\ep$ for 
$\ln(25)$ efoldings. This was repeated for different values of $\ep$.
Throughout the evolution 
we found that the Q-ball charge (defined by the Noether charge 
\mbox{$-4\pi i\int dr\;r^2\left[ \p^* \p ' - \left(\p '\right)^* \p \right]$} where r is the 
comoving radial coordinate) was conserved. So the expansion did not lead to leakage of charge. 

\begin{figure}[t]
\includegraphics[scale=0.38,angle=0]{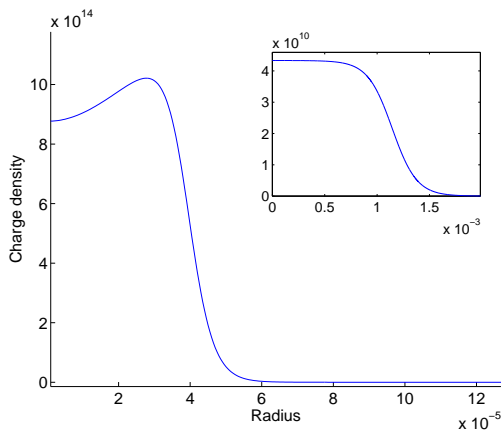}
\caption{Figure showing the charge density radial dependence for a profile in an 
expanding background (\ep=0.25), with the inset showing a typical static background profile. Note how 
the expansion causes the charge to move towards the surface of the Q-ball, compared to the static case.}
\label{chargedensity}
\end{figure}

Numerically the internal rotation frequency (in comoving units) was defined as 
\be
\om (r) \equiv Im \left[ \frac{\p ' (r)}{\p (r)} \right] 
\ee
 and we observed that for slow expansion rates its value at the core of the Q-ball  
exhibited oscillatory behaviour about a general evolution where it remained proportional to $a$ as shown in 
Fig. \ref{omega}.
Similarly, the profile amplitude $\p$ also uniformally increased as $a$ whereas the 
Q-ball co-moving radius $R$ decreased proportionally to $a^{-1}$. Recalling the definition of $\p=\vp a$, we 
see that the physical Q-ball 
parameters (radius and amplitude $\vp$) oscillate about constant values. 
The main deviation from a flat space Q-ball 
arose through the non-homogeneity of $\om (r)$, which emerged as $\ep$ increased.
Recalling that for homogeneous rotation frequencies the charge distribution is proportional to the 
amplitude of the profile function, and that no major change to the profile shape was observed, such 
a non-homogeneous $\om$ leads to a new charge density distribution.
This is evident in Fig.  \ref{chargedensity}  which shows how the expansion (with $\ep=0.25$) causes 
charge to move towards the 
surface of the Q-ball as opposed to the case of a typical charge distribution in a static background 
indicated by the inset figure shown.
However, once established, the  charge density profile did not vary with the expansion. The oscillations of 
the Q-ball parameters are to be expected in an expanding background; they correspond to oscillations 
about the minimum energy 
configuration of the Q-ball 
and are driven by the expansion of the universe. To see this, recall that the expansion leads to  time-varying 
potentials,  in which the minimum is varying with time, whilst the field is oscillating about the  minimum.

\begin{figure}[t]
\includegraphics[scale=0.38,angle=0]{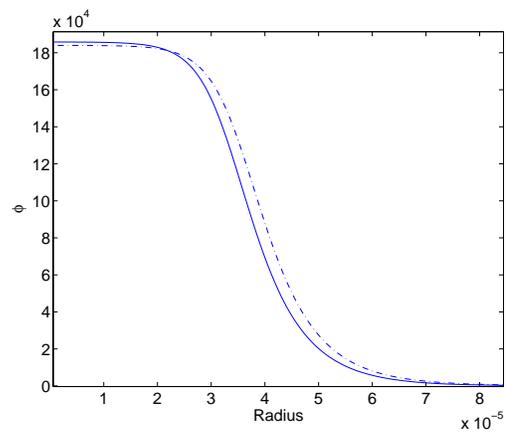}
\caption{Figure showing the numerical end profile (solid line) and the approximated profile (dotted line) 
obtained from the static solution for the case $\ep=0.1$.}
\label{match}
\end{figure}

\section{Analytic approach} \label{analytic}    

The usefulness of adopting conformal time emerges because (\ref{expandingeqnofmotion}) looks like 
(\ref{flateqnofmotion}) but with a time varying potential. It is natural, therefore, to
make the approximation that a Q-ball profile at a particular time is given by solving
(\ref{expandingeqnofmotion}) at that time,
$\p=e^{i\om \eta}f(r)$, leading to
\be
\Del^2f + \frac{\del}{\del f}\left( \half\om^2f^2 + \half\frac{a''}{a}f^2 - 
\half a^4 V(f)\right) = 0. \label{expandingansatz}
\ee
This equation is of the same form as the flat space equation (\ref{flateqnofmotion}),  but 
the price we pay is the introduction of  a new time dependent  effective potential. At a particular moment 
in time,  equation 
(\ref{expandingansatz}) has soliton solutions with the same constraints as the 
flat space case, but these are Q-ball solutions in a dynamic background. This is an 
appealing result; we can think of Q-balls in a dynamic background as flat space 
Q-balls with a time varying potential. 
It is clear that our approximate ansatz is not an exact solution as the amplitude has no 
time dependence to match the evolving time dependent potential.  Solving for 
static profiles at 
different values of $a$ and $a''$ provides an approximation for the evolving profile. 
Equation (\ref{eqofmotintime}) approaches the flat space case in the 
limit $\frac{H}{\om}\ra 0$. Therefore we parameterise our approximation by the adiabatic parameter 
$\ep=\frac{H}{\om_0}$ where $H$ is the Hubble parameter and 
$\om_0$ is the frequency in the core of the Q-ball at the initial scalefactor value. We now show 
that we can closely predict the evolving profile of the true solution, based on the corresponding static 
solution. And that deviations
from the static solutions are characterised by $\epsilon$. 

\begin{figure}[t]
\includegraphics[scale=0.38,angle=0]{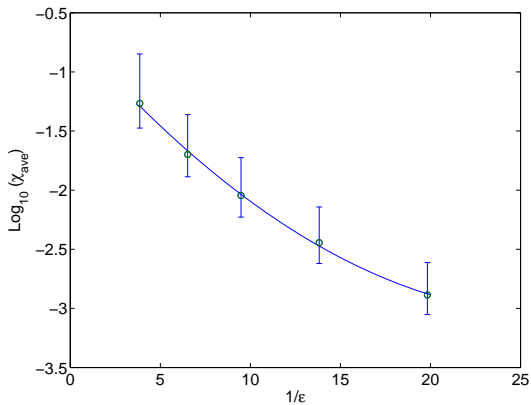}
\caption{Figure showing $\ln \chi_{\rm ave}$ as a function of $\frac{1}{\ep}$.}
\label{chiaverage}
\end{figure}

We would like to establish how in an expanding background the deviations observed from the static profiles, 
depend on $\epsilon$.
It is straightforward to solve for a static profile with the same overall charge as the exact solution and at 
the corresponding 
values of $a$ and $a''$. 
Having done this we obtain an adiabatic approximation to the evolved profile, which can be compared with the 
true solution. 
As an example of the strength of the approximation for say $\ep=0.1$,  Fig.  \ref{match} shows an overlap of 
the true 
final profile (solid line) and the corresponding `static' approximation (dotted line); they show good agreement. 
To quantify the 
accuracy of the approximation we introduce the parameter 
\be
\chi = 1 - 
\frac{\left|\int\p_{A}^{*}\p_{N}\right|^2}{\int\left|\p_{A}\right|^2 
\int\left|\p_{N}\right|^2} \geq 0,
\ee      
where $\p_{N}(r)$ is the numerically evolved profile and $\p_{A}(r)$ is the 
adiabatic approximation. 
Note that $\chi$ measures how the profiles differ including the phases of the profiles, so such measurements 
would include 
effects from non-homogeneous rotation frequencies. 
Calculating $\chi$ throughout the evolution we find that it exhibits oscillations 
of the same nature as those found in the Q-ball parameters. 
Averaging over these oscillations gives us a parameter $\chi_{ave}$ as a
measure of how well the adiabatic approximation holds as a function of $\ep$. 
Figure \ref{chiaverage} shows a plot of $\ln \chi_{\rm ave}$ against 
$\frac{1}{\ep}$, with error bars corresponding to the standard deviation of the 
oscillations from the mean (note that these are not errors in the value of $\chi_{ave}$ but are there 
to show the scale of the oscillations). We see that there is a negative logarithmic 
correlation meaning that the adiabatic approximation rapidly improves with 
decreasing $\ep$. 

Figure \ref{chiaverage} helps us to understand when we should consider expansion effects in Q-ball 
dynamics and how they vary 
with $\epsilon$ in that it provides direct information on how the expansion affects the profile. For 
example, for expansion rates 
of $\epsilon = 0.2$ we already have deviations in the profiles corresponding to around 10\%, whereas 
for $\epsilon = 0.05$ these 
reduce to a negligible value of 0.1\%. 
There is another useful non-trivial aspect to Figure 4. We, of course, expect that as $\ep \rightarrow 0$, 
we recover the static solution.
 However, an important question is how rapidly do we approach it? We can see that the deviation from the 
static Q-balls depend on high 
powers of $\ep$. Such a strong, highly non-linear, dependence is difficult is derive analytically and it 
provides important information 
in a cosmological setting where $\ep$ is varying.  
We have shown that as long as the parameter $\ep$ remains small enough, then there is a good adiabatic 
approximation describing the evolution of a Q-ball in an expanding background. In fact from the definition 
$\ep = H/\om_0$, we see that this decreases with time in any universe with 
dominating fluid whose equation of state 
$\om_{\rm dom}>-1$ (note this is not the same $\om$ as defining the Q-ball). Hence in a matter 
and radiation dominated universe the adiabatic approximation improves with time.

\section{The approximation and its limits} \label{implications}
Using the change of variables we have introduced it is possible to analytically consider the behaviour 
of a Q-ball in a slowly expanding background. Let us analyse the approximations we have made in more 
detail. Consider the general potential given in 
equations (\ref{expandingansatz}) and (\ref{potential}). This leads to an effective potential given by
\be
V_{eff}=\half\left( -\alpha a^2 +\frac{a''}{a} + \om^2 \right)\p^2 + 
\half\beta\p^4 - \half\frac{\gamma}{a^2}\p^6.\label{veff}
\ee
An adiabatic evolution allows us to solve for Q-ball properties using (\ref{veff}) and treating the 
scalefactor as a constant.  
Recalling the constraints arising in equation (\ref{omegaconstraints}) for a static flat space solution, 
we can obtain instantaneous constraints for the expanding Universe case:
\be
\al a^2 - \frac{\beta^2a^2}{4\ga} - \frac{a''}{a} \equiv \om_{\rm min}^2 \leq \om^2 \leq \om_{\rm max}^2 
\equiv \al a^2 - 
\frac{a''}{a}. \label{expandingwconditions}
\ee
We can also derive the scalefactor dependence of the Q-ball parameters. For example,
using standard thin wall limit arguments \cite{kn:coleman77}
we have the expression for the core field amplitude 
\be
|\p_0|^2 \ra  a^2\frac{\beta}{2\ga}\label{phipredict}
\ee
and for the Q-ball radius
\be
R \sim
a\frac{\beta}{2\sqrt{\ga}}\frac{1}{\Delta\om^2}+\frac{1}{\sqrt{\Delta\om^2}}.\label{Rpredict}
\ee
where $\Delta\om^2=\om^2-\om_{\rm min}^2$. 
By introducing the dimensionless parameter 
\be
\zeta=\frac{4\gamma}{\beta^2 a^2}\Delta\omega^2,
\ee
with the range $0<\zeta\leq 1$ we can write this as
\be
R\sim 
\frac{2\sqrt{\gamma}}{a\beta}\left(\frac{1}{\zeta}+\frac{1}{\sqrt{\zeta}}\right).\label{RpredictA}
\ee
We see that these expressions, derived by 
considering the equations at a particular time, correctly predict the dependence 
of the parameters on the scalefactor. Similar calculations can be used to 
predict the background dependence of other, more complicated parameters. It is encouraging that 
the analytic estimates arising from the use of the adiabatic parameter $\ep$, and the thin wall 
approximation lead to the correct values for the actual Q-ball parameters. Of course, as $\ep$ 
increases, the validity of the analytic result comes into question, eventually breaking down 
for large enough expansion rates. Fortunately though, it appears to hold well for a wide range 
of cosmologically interesting parameters. The breakdown of the approximation can be seen through 
the inhomogeneous charge distribution which emerges as $\ep$ increases as seen in figure (\ref{chargedensity}). 
These type of Q-balls are very interesting in themselves as they differ from the Q-balls discussed in 
\cite{kn:coleman} with a homogeneous rotation frequency distribution. In fact for significant expansion rates, 
they can evade the bounds imposed in 
(\ref{expandingwconditions}), which correspond to (\ref{omegaconstraints}). 
However, the bounds 
of (\ref{expandingwconditions}) do give us background dependent constraints on
the existence of nearly - homogeneous-$\om$ Q-balls. 

\section{Conclusions}\label{conc}
In this paper we have analysed Q-ball solutions in an expanding background. As expected, the presence of 
expansion changes the solution from the traditional form obtained in a Minkowski background, namely those 
which are spherically symmetric and have a  constant homogeneous frequency of rotation. However, by solving 
the system in conformal time, we can rewrite the equations of motion for the Q-ball in an expanding background 
in such a way that it mimics the case of the static background, the only difference being that the effective 
potential for the Q-ball given in equation (\ref{veff}) becomes explicitly time dependent as a result of 
the scale factor. This then allows us to introduce an adiabatic parameter $\ep=\frac{H}{\om_0}$ with 
which to analyse solutions to the full equations of motion. The smaller $\ep$ is, the closer the evolution 
is to the original static spacetime case. 
Explicit calculations of the deviations from static profiles, seen in Fig. \ref{chiaverage}, give us their 
dependence on $\ep$. This is 
important information when considering Q-balls in a cosmological context. Recalling that $\ep$ is a function 
of both the expansion rate 
and the rotation frequency we see that in backgrounds where these effects play a role, even small variations 
in the frequency or 
the expansion rate can cause movement of charge.   
We numerically showed that Q-balls existing in an expanding background conserved their charge, maintained a 
constant radius, whilst exhibiting oscillations in their frequency, a reflection of the dependence on the 
expansion of the universe. The analytic approximations we developed are able to recover the background 
dependence of these features as seen through equations (\ref{phipredict}) and (\ref{RpredictA}). We are also 
able to place background dependent constraints on the existence of Q-balls with homogeneous rotation 
frequencies(\ref{expandingwconditions}).   

We have noticed in the simulations that the Q-ball solutions, whilst still existing in that they conserve 
their charge, now develop an inhomogeneous charge distribution (see figure (\ref{chargedensity})). 
Further study of the configurations could include predicting the general form of the charge density profile 
as a function of the background expansion. Given that the Q-ball charge profile is dependent  on the expansion 
rate in a way which drives charge to the surface, we would anticipate that for large enough $\ep$, the 
charge is confined to the surface of the Q-ball. 
Such a configuration is unstable due to a large surface tension (defined by a positive energy-charge relation) 
and no bulk tension to hold it together. A similar situation for the case of non-topological cosmic 
strings is discussed in \cite{copeland} and the same conclusions are reached. 
Any cosmological considerations of Q-balls where the oscillation frequency is of the order of the Hubble 
parameter should include discussions of stability against this kind of effect. This is particularly 
important in the early universe where we are dealing with rapid expansion rates.  

\acknowledgements The work of PMS and EP was supported by PPARC.

\end{document}